\documentclass[prc,aps,onecolumn,showpacs,preprint]{revtex4}
\usepackage{graphicx}
\usepackage{amsmath}

\begin{document}

\title{Resonance parameters from $K$ matrix and $T$ matrix poles}

\author{
R.\ L.\ Workman, R.\ A.\ Arndt and M.\ W.\ Paris}
\affiliation{
Center for Nuclear Studies,
Department of Physics\\
The George Washington University,
Washington, D.C. 20052}

\date{\today}

\begin{abstract}

We extract $K$ matrix poles from our fits to 
elastic pion-nucleon scattering and eta-nucleon production
data in order to test a recently proposed method for the
determination of resonance properties, based on
the trace of the $K$ matrix. We have considered issues
associated with the separation of background
and resonance contributions, the correspondence between 
$K$ matrix and $T$ matrix poles, and the complicated behavior of eigenphases.

\end{abstract}

\pacs{PACS numbers: 11.55.Bq, 11.80.Et, 11.80.Gw }

\maketitle

In a study by Ceci and collaborators~\cite{ceci}, a method for
resonance parameter extraction was proposed, based on properties of the
trace of the $T$ matrix and the associated $K$ matrix, from a multi-channel
fit to scattering data. The relevant relations are~\cite{ceci},
\begin{equation}
\label{eqn:1}
{\rm Tr}(K) = \frac{\tilde \Gamma_R /2}{E_R - E} + \sum_{j \ne R }^N \tan \delta_j ,
\end{equation} 
and 
\begin{equation}
\label{eqn:2}
{\rm Tr}(T) = \frac{\tilde \Gamma_R /2}{E_R - E - i\tilde \Gamma_R /2} + \sum_{j \ne R}^{N}
e^{i\delta_j} \sin \delta_j ,
\end{equation}
where 
\begin{align}
\label{eqn:2.5}
\tilde \Gamma_R /2 \; = \; \Gamma_R /2 \; + (E_R - E) \tan {\delta_B},
\end{align}
$E_R$ is the resonance energy, $\Gamma_R$ is the full width, and ${\delta_B}$
is a background phase, and $N$ is the number of included channels. 
The index $R$ labels the $j=R$ element of the diagonal
$K$ matrix, and $\delta_j$ is an eigenphase. These expressions follow
directly from the expressions for the $K$ and $T$ matrices in terms of
the eigenphases
\begin{align}
\mbox{Tr}(K) &= \sum_{j=1}^N \tan\delta_j, & 
\mbox{Tr}(T) &= \sum_{j=1}^N e^{i\delta_j}\sin\delta_j,
\end{align}
assuming a single diagonal element of the $K$ matrix has the resonant form
\begin{equation}
\label{eqn:3}
\tan \delta_R = \frac{\Gamma_R /2}{E_R -E} + \tan {\delta_B} ,
\end{equation}
with a $K$ matrix pole at $E=E_R$ and the non-pole behavior collectively described
by the background phase. The quantities $\Gamma_R$, $E_R$ and $\delta_B$ are
considered as functions of the energy, $E$.

The importance of Eqs.\eqref{eqn:1} and \eqref{eqn:2}
for Ref.\cite{ceci} is that while the position of the $T$ matrix pole of the
first term in Eq.\eqref{eqn:2}, $E_R - i\tilde\Gamma_R/2$, and its residue,
$\tilde\Gamma_R/2$, depend on $\delta_B$, the
position of the $K$ matrix pole in Eq.\eqref{eqn:1}, $E_R$, and its residue,
$\Gamma_R$, do not. This is the model independence\cite{davidson} cited in
Ref.\cite{ceci}. In light of this, the authors 
of Ref.\cite{ceci} suggest the use of $K$ matrix pole positions and residues
to give a model-independent characterization of resonance structure.
Their method involves the determination of the $K$ matrix, from a given
$T$ matrix, from which the pole positions, $E_R$, and their residues, 
$\Gamma_R$, are extracted. 

We have explicitly tested this method with a 
set of amplitudes determined in recent fits to pion-nucleon scattering and
eta-nucleon production data. Using the $T$ matrix in a given partial wave,
determined in fits to the observed data \cite{sp06}, we determine the $K$
matrix, from which we extract the pole positions and residues for real
energies. This analysis yields at least two results which undermine
the utility of using the positions and residues of $K$ matrix poles
as model-independent characterizations of resonance structure. 
First, we show that assuming a different
form for Eq.\eqref{eqn:3} alters the finding of Ref.\cite{ceci} that
the $K$ matrix pole and residue are independent of the background, $\tan\delta_B$.
Second, using the $T$ matrix determined in Ref.\cite{sp06},
we numerically calculate the related $K$ matrix (see Eq.\eqref{eqn:4}) and
find that there are poles in the $T$ matrix which have no nearby
poles in the $K$ matrix for real energies. This would seem to obviate the
use of $K$ matrix poles in characterizing resonance structures observed 
in scattering experiments, since the structures present in the $T$ matrix
do not necessarily appear in the $K$ matrix. 

Prior to describing our numerical results, we revisit the derivation of
Eqs.\eqref{eqn:1} and \eqref{eqn:2} in relation to the assumptions of
Eq.\eqref{eqn:3}. We then compare the result of Ref.\cite{ceci} with the result 
obtained with a different assumption (following Dalitz\cite{dalitz})
for the parameterization\cite{davidson2} of the resonant eigenphase of Eq.\eqref{eqn:3}.

The $K$ matrix we use is real for energies above all thresholds
considered in the problem, and is related to the $T$ matrix by
\begin{equation}
\label{eqn:4}
T \; = \; K \; ( 1 - i K )^{-1} .
\end{equation}
The real symmetric $K$ matrix is diagonalized by an orthogonal
transformation, $U$ as
\begin{equation}
K_D = U^T\; K \; U .
\end{equation}
This matrix also diagonalizes the $T$ matrix, and therefore the $S$ matrix,
defined as $S = 1 + 2iT$. Since $S$ is a unitary matrix,
\begin{equation}
( S_D )_{ij} \; = \; U^T S U \; = \; \delta_{ij} \; e^{2 i \delta_i} 
\end{equation}
where the $\delta_i$ are eigenphases. Using the relation between $K$
and $T$ matrices above, we have
\begin{equation}
\label{eqn:K_D}
( K_D )_{ij} \; = \; i (1 - S_D ) (1 + S_D )^{-1} \; = \; \delta_{ij} \tan \delta_i .
\end{equation}  
Having determined the eigenphases, we can reconstruct the physical
$T$ matrix
\begin{equation}
T_{if} \; = \; ( U T_D U^T )_{if} = \sum_{\alpha} 
U_{i \alpha} U_{f \alpha} e^{i\delta_{\alpha}} \sin \delta_{\alpha} .
\end{equation}
Taking the trace of Eqs.~(9) and (10) gives the result in Eq.~(4).

We first examine the use of these relations in a simple scenario
including a single resonant eigenphase and neglect any background
effects. Assuming only one eigenphase $(\delta_R )$ passes through 
$\pi / 2$ at energy $E_R$ and neglecting others, the resonant eigenphase
must have the form
\begin{equation}
\label{eqn:delta_R}
\tan \delta_R \; = \; \frac{\Gamma_R / 2}{ E_R - E },
\end{equation}
which leads to 
\begin{equation}
\label{eqn:toyT}
T_{i f} \; = 
\; \frac{1}{2} \frac{ \Gamma^{1/2}_i \Gamma^{1/2}_f }{E_R - E -i \Gamma_R/2}.
\end{equation}
with $\Gamma_i = U_{iR}^2 \Gamma_R $ and $\sum_i U_{iR}^2 = 1$ 
(orthogonality) giving $\sum_i \Gamma_i = \Gamma_R$. 
The result, Eq.\eqref{eqn:toyT} is consistent with Eqs.\eqref{eqn:1}
and \eqref{eqn:2} for $\delta_B = 0$. Next we consider how background
can be added, and whether a single dominant eigenphase is appropriate. 
These questions have been addressed in the works of Dalitz~\cite{dalitz}, 
Goebel and McVoy~\cite{mcv}, and Weidenm\"uller~\cite{weiden}.

Consider first the addition of a background phase to Eq.\eqref{eqn:delta_R}. 
One way of doing this is the {\em ans\"{a}tz} of Eq.\eqref{eqn:3}
employed in Ref.\cite{ceci} (see also Ref.\cite{martin}) and used to
obtain Eqs.\eqref{eqn:1} and \eqref{eqn:2}. This leads to the
model independence of Ref.\cite{ceci} described above. An alternative 
parameterization of the resonant eigenphase is considered in 
Refs.\cite{dalitz,mcv,weiden}. The {\em ans\"{a}tz} used there also 
assumes a single dominant eigenphase, which rises through $\pi /2$, but 
posits a phase-addition rule: the resonant eigenphase, $\tilde\delta_R$
has the form
\begin{align}
\label{eqn:dalitz}
\tilde\delta_R &= \tilde\delta_B + \delta_P,
\end{align}
where $\tilde\delta_B$ is the background phase which determines the 
eigenphase far from the resonance energy, and $\delta_P$ is the resonant 
(pole) contribution. This form of resonance and background separation
modifies the above conclusion of model independence. 
We consider the phase-addition rule in some detail 
to clarify this point.

As a function with a simple pole, the resonant contribution, $\delta_P$ 
may be written in general as
\begin{align}
\label{eqn:delP}
\tan\delta_P &= \frac{\gamma(E)/2}{E^*(E)-E},
\end{align}
where the position of the pole is given by $E^*_P$ $(E^*(E^*_P)-E^*_P=0)$
and the function $\gamma(E)$ goes to a non-zero constant at the pole. Note that,
far from the pole, the eigenphase shift $\tilde\delta_R$
reduces to the non-pole part, $\tilde\delta_B$. Using Eqs.\eqref{eqn:dalitz} 
and \eqref{eqn:delP} we compute the resonant element of the diagonal $K$
matrix as
\begin{align}
\tan\tilde\delta_R &= 
\frac{\frac{1}{2}\gamma+(E^*-E)\tan\tilde\delta_B}
     {(E^*-E)-\frac{1}{2}\gamma\tan\tilde\delta_B},
\end{align}
which leads to a $K$ matrix with the trace
\begin{align}
\mbox{Tr}(K) &= \frac{\overline{\Gamma}(E)/2}{\overline{E}^*_P(E)-E}
+ \sum_{j\ne R}^N \tan\delta_j,
\end{align}
where $\overline{\Gamma}(E)/2 = \gamma/2+(E^*-E)\tan\tilde\delta_B$ and
the location of the pole in $\mbox{Tr}(K)$ is $\overline{E}^*_P$, where
\begin{align}
\left.
[(E^*(E) - E)-\frac{\gamma(E)}{2}\tan\tilde\delta_B(E)]
\right|_{E=\overline{E}^*_P}&=0,
\end{align}
In general, $E^*_P\ne \overline{E}^*_P$ and the pole position of the $K$
matrix, $\overline{E}^*_P$ depends on the background, $\tan\tilde\delta_B$.
The residue also depends on $\tilde\delta_B$ since
$\overline{\Gamma}(\overline{E}^*_P)=\gamma/\cos^2\tilde\delta_B$.

We could anticipate this result by comparing the forms 
Eqs.\eqref{eqn:3} and \eqref{eqn:dalitz}. In Eq.\eqref{eqn:3}, used in
Ref.\cite{ceci}, the location of the $K$ matrix pole, $E_R$ is
independent of $\delta_B$. The resonant structure,
$\Gamma/[2(E_R-E)]$, and the non-resonant contribution are {\em assumed} 
to be decoupled. That is, if $\tan\delta_B$ is a bounded function
of the energy, its value cannot affect the energy where the resonant
eigenphase, $\delta_R$ is $\pi/2$. 
In the ``Dalitz form,'' Eq.\eqref{eqn:dalitz},
the location of the pole in the $K$ matrix, determined by the 
energy $E_R$ where the phase $\delta_R \to \pi/2$, is affected
by the ``background phase,'' $\delta_B$. Since the true form of the $K$ 
matrix is unknown, the existence of alternative forms complicates the
extraction of pole positions.
In fact, in dynamical models of scattering amplitudes,
the location of the $K$ matrix pole is expected to depend, perhaps strongly,
on the non-resonant (or background) contribution to the amplitude
\cite{Paris:2009inprep}.

Turning to the $T$ matrix, in place of Eq.\eqref{eqn:toyT}, the result is
\begin{equation}
T_{i f} \; = \; \frac{1}{2} e^{2i\tilde\delta_B} 
\frac{ \Gamma'^{1/2}_i \Gamma'^{1/2}_j }{E'_R - E -i \Gamma' / 2}
+ U_{iR} U_{jR} e^{i\tilde\delta_B} \sin \tilde\delta_B,
\end{equation}
for the corresponding $T$ matrix element with resonance `mass' and `width' 
shifted from the $K$ matrix pole parameters. Thus, a different scheme for the 
addition of resonance and background contributions can alter the relationship 
between $K$ matrix and $T$ matrix resonance masses. 

\begin{figure}
\includegraphics[ width=350pt, keepaspectratio, clip]{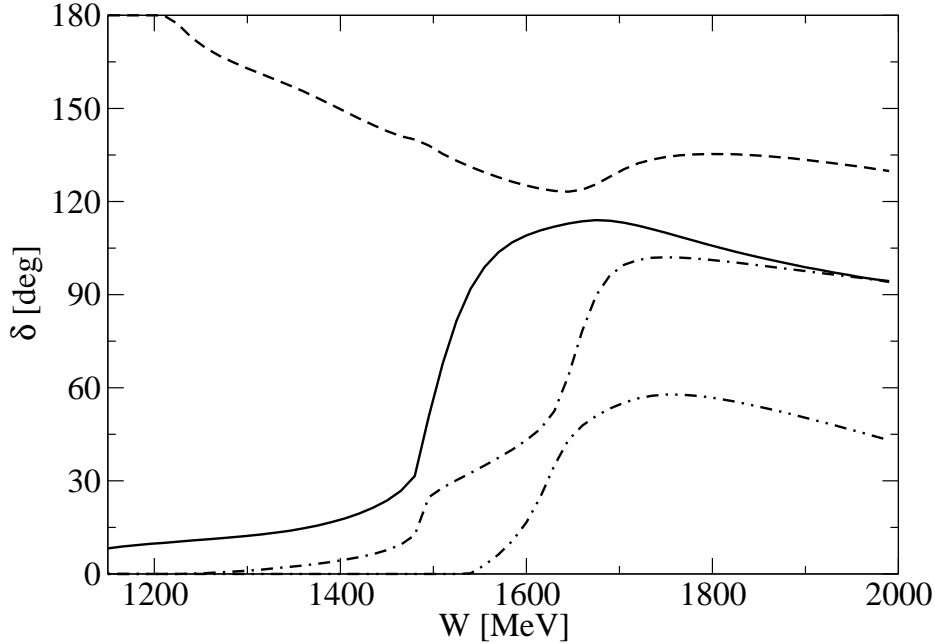}
\caption{\label{fig:s11egnph}The eigenphases in a four-channel fit to the
$S_{11}$ partial wave from the {\sc SP06} solution from SAID.}
\end{figure}

As another example of the model dependence of $K$ and $T$ matrix poles, and
to address the question of whether a single resonant eigenphase is appropriate,
we consider the following simple $S$ matrix from Refs.\cite{mcv,weiden}
\begin{equation}
\label{eqn:Sij}
S_{ij} \; = \; e^{i ( \phi_i + \phi_j ) }
\left[ \delta_{ij} 
+ i \frac{ {\Gamma_i}^{1/2} {\Gamma_j}^{1/2} }{ E_R - E -i\Gamma /2} \right],
\end{equation}
to show the effect of background on eigenphases. This 
$S$ matrix is symmetric and far from the resonance energy is diagonal 
(the elastic background approximation) with eigenphases $\phi_i$\cite{fn1,fn2}. 
Applying the unitary 
transformation diagonalizing Eq.\eqref{eqn:Sij} and taking the determinant, 
yields
\begin{equation}
\label{eqn:weid}
e^{2 i \sum_i \delta_i} \; = \; e^{2 i \sum_i \phi_i} \; 
           \frac{E_R - E + i\Gamma /2}{E_R - E -i\Gamma /2},
\end{equation}
where $\delta_i$ is an eigenphase and the last factor has the phase 
behavior of an elastic resonance at $E_R$. From Eq.\eqref{eqn:weid}
we see the above phase-addition rule, Eq.\eqref{eqn:dalitz},
if only a single eigenphase is significant. In general, however, it is
the sum of eigenphases that displays resonance behavior.  

A few further comments on the parameterization of eigenphases may be useful.
Weidenm\"uller~\cite{weiden} has shown that individual eigenphases have an
energy dependence determined largely by the background. Through an application
of Wigner's no-crossing theorem, he finds no single eigenphase increasing by
$\pi $, except for special values of the background phases. As a result, the
eigenphases `repel' rather than crossing, the $N$ eigenphases individually
increasing only by an average of $\pi /N$ over the width of the resonance in
some cases. An example of this behavior is given in Fig.\eqref{fig:s11egnph},
which shows the eigenphases calculated from {\sc SAID} \cite{sp06} for the 
$S_{11}$ partial wave, containing two resonances.

Goebel and McVoy have applied the eigenvalue method to resonant $d - \alpha$
scattering~\cite{mcv} data to explicitly study this behavior. Eigenvalues for
this two-channel scattering matrix were also given, showing the
appearance of square-root branch points which complicate the energy
dependence~\cite{gold}. There is a cancellation occurring when the sum of
eigenvalues or eigenphases is taken, and this supports the basic idea of using
a trace, as proposed in Ref.~\cite{ceci}. A direct relation between resonance
energy and the sum of eigenphases is given by the equation~\cite{smith,arndt}
\begin{equation}
{\rm tr} \; Q \; = \; 2 \hbar \sum_i \frac{d \delta_i}{d E},
\end{equation}
relating the trace of Smith's time-delay matrix to the energy
derivative of the sum of eigenphases, $\delta_i$. One diagonal element
of the $Q$ matrix has recently been shown to correlate precisely 
with the T-matrix pole positions of resonances~\cite{arndt}.

\begin{table}[t]
\begin{tabular*}{0.75\textwidth}{@{\extracolsep{\fill}}c|cc|cc}
$\ell_{JT}$ & \multicolumn{2}{c}{$T$ poles} &\multicolumn{2}{c}{$K$ poles} \\
\hline
$S_{11}$ & $(1500,50)$ & $(1650,40)$ & $1535$ & $1675$ \\
$P_{11}$ & $(1360,80)$ & $(1390,80)^\dag$ & $-$ & $-$ \\
$P_{13}$ & $(1665,175)$ & & $-$ &  \\
$D_{13}$ & $(1515,55)$ & & $-$ &  \\
$D_{15}$ & $(1655,70)$ & & $1760$ &  \\
$F_{15}$ & $(1675,60)$ & $(1780,130)$ & $-$ & $-$ \\
\end{tabular*}
\caption{\label{tab:pole1}Pole positions in complex energy plane of 
$T$ and $K$ matrix for the $\pi N\to\pi N$ reaction
from {\sc SAID} \cite{sp06} for isospin $T=\tfrac{1}{2}$ partial
waves. Each $T$ pole position is expressed in terms of its real and imaginary 
parts ($M_R,-\Gamma_R/2$) in MeV. Only $K$ matrix pole positions which satisfy
$1.1 \mbox{ GeV} < W < 2.0 \mbox{ GeV}$ are considered. $^\dag$This pole is
located on the second Riemann sheet.}
\end{table}

To explicitly test the method of Ref.~\cite{ceci}, we have taken amplitudes
determined from our fits to pion-nucleon elastic scattering data \cite{sp06},
and the reaction $\pi N\to \eta N$.  The parameterization we use is based on 
the Chew-Mandelstam (CM) $K$ matrix, which builds in cuts associated with the
opening of $\eta N$, $\pi \Delta$, and $\rho N$ channels. The CM form is analytic
and generates a $T$ matrix which is unitary and can be continued into the
complex plane to find poles on the various sheets associated with the $\eta N$,
$\pi \Delta$, and $\rho N$ channels. The fits can, in principle, include
couplings to any of the channels, though the $\pi \Delta$ and $\rho N$ channels
are not constrained by data.  However, the amplitude associated with each
channel has, by construction, the proper threshold behavior, cuts, and pole
positions. Amplitudes in the elastic channel are further constrained by forward
and fixed-$t$ dispersion relations.

\begin{table}[b]
\begin{tabular*}{0.75\textwidth}{@{\extracolsep{\fill}}c|cc|cc}
$\ell_{JT}$ & \multicolumn{2}{c}{$T$ poles} &\multicolumn{2}{c}{$K$ poles} \\
\hline
$S_{31}$ & $(1595,70)$ &              & $1660$ &        \\
$P_{31}$ & $(1770,240)$ &   & $-$  &   \\
$P_{33}$ & $(1210,50)$ & $(1460,200)$ & $1232$ & $-$ \\
$D_{33}$ & $(1630,125)$ & & $-$ &  \\
$D_{35}$ & $(2000,195)$ & & $-$ &  \\
$F_{35}$ & $(1820,125)$ & & $-$ &  \\
\end{tabular*}
\caption{\label{tab:pole3}Pole positions in complex energy plane 
as in Table \ref{tab:pole1} for isospin $T=\tfrac{3}{2}$ partial
waves.}
\end{table}

Thus far, we have implicitly assumed that there is a direct correspondence 
between $K$ matrix and $T$ matrix poles. It is known~\cite{levi} that this
is not true in general. For example, in the CM approach
it is possible to generate $T$ 
matrix poles for the resonances without explicitly adding a pole to the CM 
$K$ matrix~\cite{ford}. If, however, a pure CM $K$ matrix pole representation is used
\begin{equation}
K_{ij} \; = \; \frac{\gamma_i \gamma_j (\rho_i \rho_j)^{1/2} }{ E_K - E} ,
\end{equation} 
the resulting $T$ matrix is 
\begin{equation}
T_{ij} 
\; = \; 
\frac{\gamma_i \gamma_j (\rho_i \rho_j)^{1/2} }
{ E_K - E - \sum_n \gamma_n^2 C_n - i \sum_n \gamma_n^2 \rho_n }
\end{equation}
where $\rho_i$ is the phase space factor for the $i^{th}$ channel, and $C_i$ is the
real part of the Chew-Mandelstam function, obtained by integrating phase space factors
over appropriate unitarity cuts. 

The fit under consideration uses a 
parameterization of a CM $K$ matrix\cite{ford},
from which the unitary $T$ matrix is calculated. The $K$ matrix, defined 
by Eq.\eqref{eqn:4}, is computed from the calculated $T$ matrix
as $K=T(1+iT)^{-1}$. The resulting $K$ matrix was checked for
consistency by reproduction of the $T$ matrix from Eq.\eqref{eqn:4},
and checked for unitarity at each stage.  The $K$
matrix was then searched for poles at energies associated with well-known
resonances. When poles did appear in a given amplitude, we confirmed that they
appeared in each associated amplitude at the same energy. However, we did not
generally find poles closely associated with ($T$ matrix) resonance energies, 
nor did we find that each resonance produced a $K$ matrix pole, as shown in 
Tables \ref{tab:pole1} and \ref{tab:pole3}. 
If an explicit pole was
inserted into the CM $K$ matrix, then this approach generated a corresponding
$K$ matrix pole.  This was the case for the $\Delta (1232)$ resonance,
where we found a $K$ matrix pole at 1232 MeV.
$K$ matrix poles also
appeared near the $N(1535)$, $N(1650)$, and $\Delta (1620)$ resonance masses,
in the $\pi N$ $S_{11}$ and $S_{31}$ partial waves, though no explicit CM $K$ matrix
poles were used in the fit. For the $P_{11}$ and $D_{13}$ partial waves,
however, no CM $K$ matrix poles were used in the fit, and no $K$ matrix
poles were found near the resonance masses. In all of these cases, resonance
poles did appear in the corresponding $T$ matrices.

In conclusion, we have not found a simple association between $K$ matrix and
$T$ matrix poles for use in the extraction of resonance properties. We have
argued that: {\em i)} $K$ matrix poles are not generally independent
of background contributions, {\em ii)} a pole in the $T$ matrix does not
necessarily imply a pole in the $K$ matrix. Therefore, $K$ matrix poles do
not appear to be useful candidates for characterizing resonance parameters
obtained from scattering amplitudes. Applied to a particular $S$ matrix 
obtained from a fit to pion-nucleon and eta-nucleon scattering 
data \cite{sp06} we find no one-to-one association between $K$ matrix and 
$T$ matrix poles. We have also noted that the separation of background and 
resonance contributions is not unique and that eigenphase behavior may be 
more complicated than the form chosen in Ref.~\cite{ceci}. We have noted an 
explicit counterexample for the parameterization of the resonant eigenphase, 
specifically Eq.\eqref{eqn:dalitz}, which violates
the model independence of Ref.\cite{ceci}.  We are currently exploring the
behavior of eigenphases using $S$ matrices from scattering amplitudes in order
to determine whether eigenphase repulsion is as common as suggested in
Ref.~\cite{mcv}.

\begin{acknowledgments}

This work was supported in part by the U.S. Department of
Energy Grant DE-FG02-99ER41110 and funding provided by Jefferson Lab. 
\end{acknowledgments}
\eject

\eject


\begin{thebibliography}{99}
\bibitem{fn1}
Note that a product form of background and resonant $S$ matrices
$S = S_B S_R$, generally yields a non-symmetric $S$ matrix.
This violates time-reversal
invariance, which is required to obtain a real $K$ matrix.
\bibitem{fn2}
The $2\times2$ $S$
matrix in Eq.(19) is sufficiently simple to allow an explicit calculation of
eigenphases~\cite{mcv}, $\sigma_{\pm}$, using
\begin{equation}
\sigma_{\pm} = \frac{1}{2} \left( S_{11} + S_{22} \right) \pm \frac{1}{2}
\left[ \left( S_{11} - S_{22} \right) ^2 + 4 S_{12}^2 \right] ^{1/2} .
\end{equation}
\end{thebibliography}
\end{document}